\documentstyle[prl,aps,epsf,psfig,multicol]{revtex}
\begin{document}

\title{Multi-Self-Overlap Ensemble for protein folding:
ground state search and thermodynamics}
\author{George Chikenji$^{1}$\footnote{chikenji@godzilla.phys.sci.osaka-u.ac.jp}, Macoto Kikuchi$^{1}$, and Yukito Iba$^{2}$ \footnote{iba@ism.ac.jp} }
\address{$^{1}$ Department of Physics, Osaka University, Toyonaka
560-0043, Japan\\
$^{2}$ The Institute of Statistical Mathmatics, Minatoku, Tokyo
106-8569, Japan}

\date{\today}

\maketitle

\begin{abstract}
Long chains of the HP lattice protein model are studied by
the Multi-Self-Overlap Ensemble(MSOE) Monte Carlo method,
which was developed recently by the authors\cite{Iba&Chike&Kiku}.
MSOE successfully finds the lowest energy states reported before
for sequences of the chain length $N=42\sim 100$ in
two and three dimensions.
Moreover, MSOE realizes the lowest energy state that ever found
in a case of $N=100$.
Finite-temperature properties of these sequences are also investigated by MSOE.
Two successive transitions are observed between the native and
random coil states.
Thermodynamic analysis suggests
that the ground state degeneracy is relevant to the order of
the transitions in the HP model.
  \vspace{4pt}
\end{abstract}

\begin{multicols}{2}

%%%%%%%%%%%%%%%%%%%%%%%%%
%    Introdunction      %
%%%%%%%%%%%%%%%%%%%%%%%%%
Protein folding\cite{Creighton} is
one of the most interesting problems in biological science.
How amino acid sequences code their native structures,
functions, and thermodynamical behavior?
To answer these questions, it is useful to study
simplified models.
Among them, lattice protein models\cite{Go,Lau&Dill}
have been playing important roles
in theoretical studies of protein folding.
For relatively short chains,
the exact enumerations of all the conformations are possible.
In dealing with longer chains, however,
we encounter difficulty,
because dynamical Monte Carlo methods are not efficient for
lattice protein models\cite{Yue}.
One of the reasons why conventional Monte Carlo
algorithms are so slow is
a rugged free-energy landscape at low temperature
due to the heterogeneity of the interactions.
For overcoming slow dynamics due to the ruggedness
in other random systems such as the spin glasses,
extended ensemble Monte Calro methods have been developped:
the multicanonical algorithm~\cite{Berg&Neuhaus,Lee},
the simulated tempering algorithm~\cite{Marinari&Parisi},
the exchange Monte Carlo algorithm\cite{Nemoto&Hukushima,Tesi}, and so on.
In case of lattice polymers, however,
topological barriers due to the self-avoiding condition
cause further difficulty,
because they are independent of energy and thus cannot be cured
by the use of the abovementioned extended ensemble methods.
In fact, a work by Shakhnovich {\it et al.}~\cite{xxx}
suggested that faster dynamics is attained
by relaxing the self-avoidingness.

Recently, we proposed a new Monte Carlo method
called {\sl Multi-Self-Overlap Ensemble (MSOE)}~\cite{Iba&Chike&Kiku}
by generalizing the idea of the multicanonical ensemble.
In this method,
the self-avoiding condition is systematically weakened.
While the resultant ensemble
contains some portions of non-physical self-overlapping configurations,
the correct canonical ensemble
of the physical conformations
can be reconstructed through the histogram reweighting procedure.
We observed a considerably fast relaxation for a simple lattice
heteropolymer
compared with the conventional multicanonical ensemble method.

In this letter, we apply MSOE to long chains of
the HP lattice protein model\cite{Lau&Dill}
in two and three dimensions.
We demonstrate that the low energy states are found successfully
and thermodynamic properties are calculated at the same time.
It should be noted that the principle of MSOE is
independent of specific choices of the interactions between monomers.
Although we restrict ourselves to the HP model
in the following,
extentions to models with other interactions,
such as the one with Miyazawa-Jernigan contact matrix~\cite{MJ},
are straightfoward.

%%%%%%%%%%%%%%%%%%%%%%%%%
%      algorithm        %
%%%%%%%%%%%%%%%%%%%%%%%%%
Let us explain MSOE algorithm briefly.
First, we
define the degree of violation of self-avoidance $V$ as
\begin{equation}
    V = \sum_{i \in G}(n_i - 1)^2
\end{equation}
where $n_i$ is the number of monomers on a lattice point $i$, and
$G$ denotes a set of lattice points which are occupied by at least one
monomer.
For self-avoiding conformations, $V = 0$.
We assume that the definition of the original energy function $E$
can be extended to conformations with self-overlaps
in a reasonable manner.
Then, we determine the weight factor $ P_{g} \propto \exp [-g(E,V)]$
through pleliminary runs
as in the case of the conventional multicanonical  
ensemble\cite{Berg&Neuhaus},
so that the bivariate histogram of $(E,V)$
is sufficiently flat ina prescribed range 
\end{multicols}
%%%%%%%%%%%%%%%%%%%%%%%%%%%%%%%%%%%%%%%%%%%%%%%%%%%%%%%%%%%%%%%%
\vglue-.2cm
\begin{table}
\caption{
The values of the lowest energy reported by several authors
for three sequences of 2D HP model.}
\begin{tabular}{cccccc}
   $N$   &    sequence          &     $E_{\rm min}$      & Ref.   \\
 \hline
     &
     &  $-37$
     &  \cite{Unger&Moult} & \\
     &
     &  $-40$
     &  \cite{Toma&Toma} &\\
        64
     &   
$H_{12}PHPHP_2H_2P_2H_2P_2HP_2H_2P_2H_2P_2HP_2H_2P_2H_2P_2HPHPH_{12}$
     &  $-42$
     &  \cite{Beutler&Dill} &\\
     &
     &  $-42$
     &  \cite{PERM}  &\\
     &
     &  $-42$
     &present study  &\\
  \hline
     &
     &  $-46$
     &  \cite{Ramakrishnan} &\\
       100
     &   
$P_3H_2P_2H_4P_2H_3(PH_2)_3H_2P_8H_6P_2H_6P_9HPH_2PH_{11}P_2H_3PH_2PHP_2HP 
H_3P_6H_3$
     &  $-49$
     &  \cite{PERM} &\\
     &
     &  $-50$
     &present study &\\
  \hline
     &
     &  $-44$
     &  \cite{Ramakrishnan} &\\
       100
     &  
$P_6HPH_2P_5H_3PH_5PH_2P_2(P_2H_2)_2PH_5PH_{10}PH_2PH_7P_{11}H_7P_2HPH_3P_ 
6HPH_2$
     &  $-47$
     &  \cite{PERM} &\\
     &
     &  $-47$
     &present study &\\
\end{tabular}

\end{table}
%%%%%%%%%%%%%%%%%%%%%%%%%%%%%%%%%%%%%%%%%%%%%%%%%%%%%%%%%%%%%%%%
\begin{multicols}{2}
$E_{min} \leq E \leq E_{max}, \, 0 \leq V \leq
V_{max}$.
We set $E_{min}$ to a value
which is definitely lower than the ground state energy, and
set $V_{max}$ to a value $N/10$ $\sim$ $N/5$.
We actually use the entropic sampling method~\cite{Lee} in this paper.
After the weight factor $g(E,V)$ is determined,
a measurement run is performed.
The canonical averages at temperature $T$ is calculated
according to the histogram reweighting formula,
\begin{equation}
  \langle A \rangle_T
=
      \frac{ \sum_i'  A(\Gamma_i) P^{-1}_g(\Gamma_i) \exp(-
E(\Gamma_i)/T ) }
           { \sum_i'  P^{-1}_g(\Gamma_i) \exp(-E(\Gamma_i)/T ) } {}_,
\end{equation}
where $\Gamma_i$ represents a conformation at the $i$th Monte Carlo step
and the summations are taken only over the self-avoiding conformations.

%%%%%%%%%%%%%%%%%%%%%%%
%  How MSOE works ?   %
%%%%%%%%%%%%%%%%%%%%%%%
How MSOE works?
In Fig.~1, observed transitions during a MSOE simulation are
plotted on the $(E,V)$ plane (This example is taken from
a simulation for the 2D $N=64$ sequence shown in Table I,
of which we will discuss below).
We can clearly see {\sl bridges} (or {\sl paths})
between self-avoiding low energy states
where self-overlapping states are used as {\sl stepping stones}.
That is, while no direct
transition is seen between $(E,0)$ and $(E',0)$,
there are paths that utilize the states with non-zero $V$
as itermediate states.
The existence of such paths are a key feature of MSOE,
which effectively facilitates the relaxation.
The idea behind MSOE can also
be extended to off lattice models.
For example,
one can consider the hard core repulsive part of the energy
as an off-lattice counterpart of the self-avoiding conditions.
Other types of bivariate extention of the multicanonical algorithm
for off-lattice protein models have also been discussed
by Higo {\it et al.}\cite{Higo}.
They, however, did not incorporate unphysical conformations
for the purpose of attaining fast relaxations.
MSOE also shares some ideas in common with
the {\sl ghost} polymer procedure by Wilding and M\"uller
for dense polymer systems\cite{Wilding&Mueller}.
\begin{figure}
    \begin{center}
    \epsfile{file=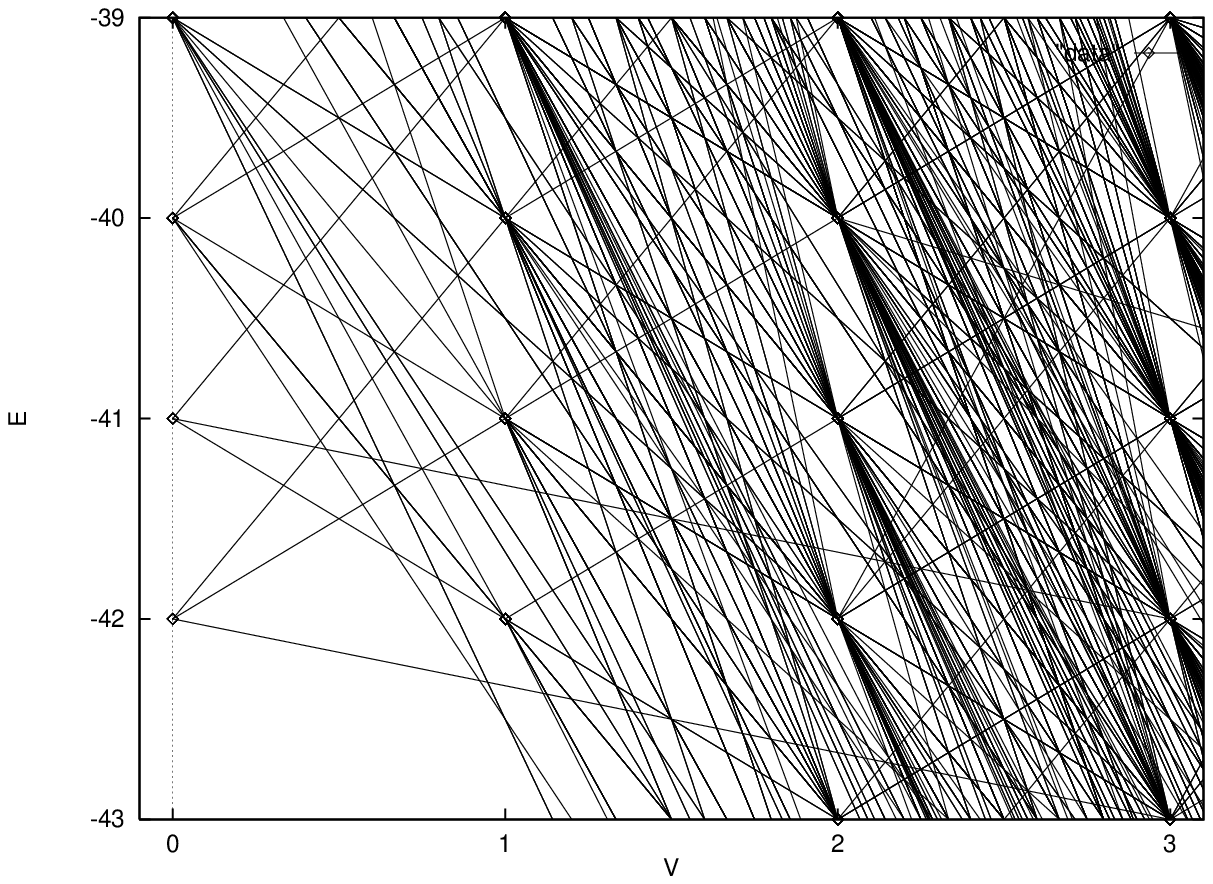,height=5.0cm}
    \end{center}
    {FIG.1.Observed transitions on the $(E,V)$ plane during a MSOE run
     of the $N=64$ HP sequence in Table I.
     Solid lines connecting two states represent the transitions between them.
     Only a small part of the entire $(E,V)$ plane near the ground state is shown.}
\label{transition}
\end{figure}

%%%%%%%%%%%%%%%%%%%%%%%
%      Model          %
%%%%%%%%%%%%%%%%%%%%%%%
Let us discuss the results of simulations
for the HP protein model\cite{Lau&Dill},
in which a protein consists of a self-avoiding chain on a lattice
with two types of amino acids: H(hydrophobic) and P(Polar).
The energy $E$ of a chain conformation is determined only
by the number of H--H contacts $h$ as
$E = - |\epsilon| h$, where $|\epsilon|$ is a positive constant
(we measure the energy in the unit of $|\epsilon|$ hereafter).
In MSOE,
we use the same definition of energy as the original one
also for self-overlapping conformations.
%%%%%%%%%%%%%%%%%
%   Dynamics    %
%%%%%%%%%%%%%%%%%
In principle, MSOE can be implemented with arbitrary dynamics.
We use the following elementary moves in this paper:
(1)Jacknife move, (2)One bead flip, (3)Pivot operator.
A description of the move (1) is given in ref.~\cite{Iba&Chike&Kiku} and
the moves (2) and (3) are illastrated in ref.\cite{Sokal}.
%%%%%%
All CPU times quote below refer to PCs with 500MHz DEC 21164A chip.

%%%%%%%%%%%%
% 2d N=64  %
%%%%%%%%%%%%
First we consider a
two dimensional HP lattice protein with the chain length $N=64$ in Table I,
several methods so far~\cite{Unger&Moult,Toma&Toma,Beutler&Dill,PERM}.
For this sequence, the ground state energy is
believed to be $E=-42$;
one of the ground state conformations is shown
in Fig.~5 of ref.~\cite{PERM}.
We calculated the weight factor
within 30 hours with $E_{min} = -60$ and $V_{max} = 10$.
Figure 2 shows the marginal distribution of $(E,V)$
obtained by the measurement run.
The marginal distribution is sufficiently flat including the state
$(-42,0)$,
and we can caluculate thermodynamic quantities
of the sequence at {\it any temperature} using it.
\begin{figure}[ht]
    \begin{center}
    \psfig{file=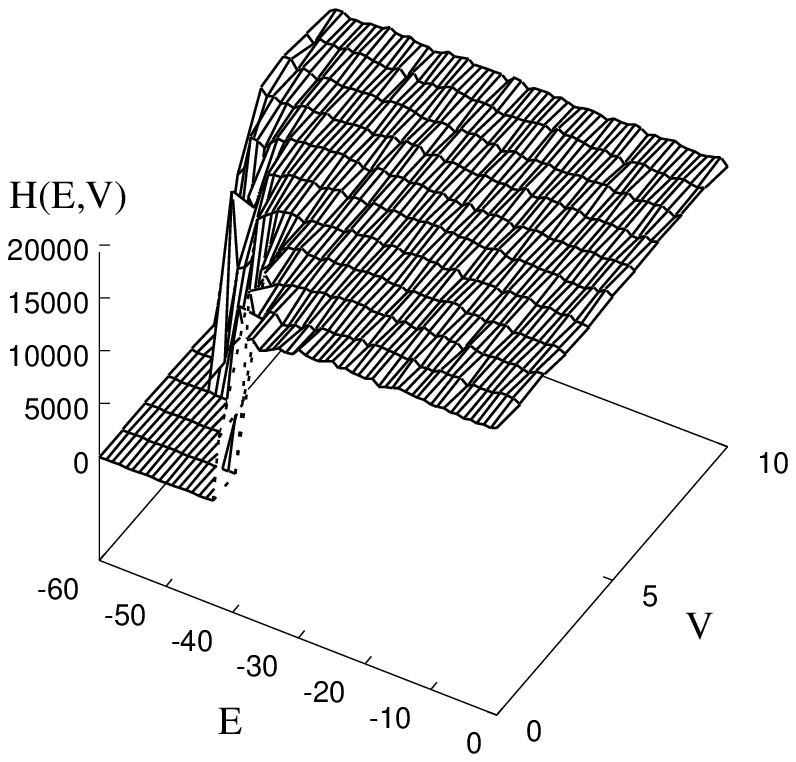,height=6.0cm}
    \end{center}
    {FIG.2. Bivariate histogram $H(E,V)$ obtained by a measurement run
     of the $N=64$ HP sequence shown in Table I.}
    \label{histogram}
\end{figure}

The values of the lowest energy found by several methods
are listed in Table I.
{\sl Core directed chain Growth method(CG)}~\cite{Beutler&Dill}
has given the ground states.
It is, however, specialized in the search of the ground state,
and thus
cannot be used for thermodynamics.
Bastolla {\it et al.} also have found these states ($E=-42$)
by {\sl pruned-enriched Rosenbluth method(PERM)}~\cite{PERM}
with an assumption that the ground states
do not contain any non-bonded HP neighbor pair
(the lowest energy obtained without this assumption
was $E=-40$).
Such an additional assumption will cause uncontrollable
biases in calculations of finite temperature properties.
Moreover, as we will see later,
there are examples in which low energy states does not
satisfy this assumption.
Then, as far as we know, MSOE is currently the only method which can
calculate finite temperature properties as well as the
low energy states of this sequence.

We calculated the specific heat and the gyration radius.
We found a peak at $T\sim 0.55$ and a shoulder
a $T\sim 0.4$ in the specific heat curve.
The shoulder at $T=0.4$ corresponds to the transition
between the ground states and compact globule states,
which is a first-order-like transition
since the energy distribution becomes bimodal.
The peak at $T=0.55$, on the other hand, is
the transition between compact globule states and the random coil;
In this case the energy distribution is unimodal, that is,
the transition is second--order--like.

%%%%%%%%%%%%
% 2d N=100 %
%%%%%%%%%%%%
Consider another sequence of $N=100$
shown in the second row of Table I.
Bastolla {\it et al.}~\cite{PERM}
have found conformations with $E = -49$
($E=-48$ without the abovementioned assumption on
the ground state conformations).
On the other hand, we found conformations
with $E = -50$ within 50 hours by MSOE
taking $E_{min} = -65$ and $V_{max} = 10$.
Figure~3 shows a typical conformation with $E = -50$.
This conformation contains
some non-bonded HP neighbor pairs.
Thus, it can never be found when these pairs are not allowed.
Unfortunately, we have not yet obtained the weight factor that gives
a sufficiently flat distribution of $(E, V)$.
\begin{figure}
    \begin{center}
    \epsfile{file=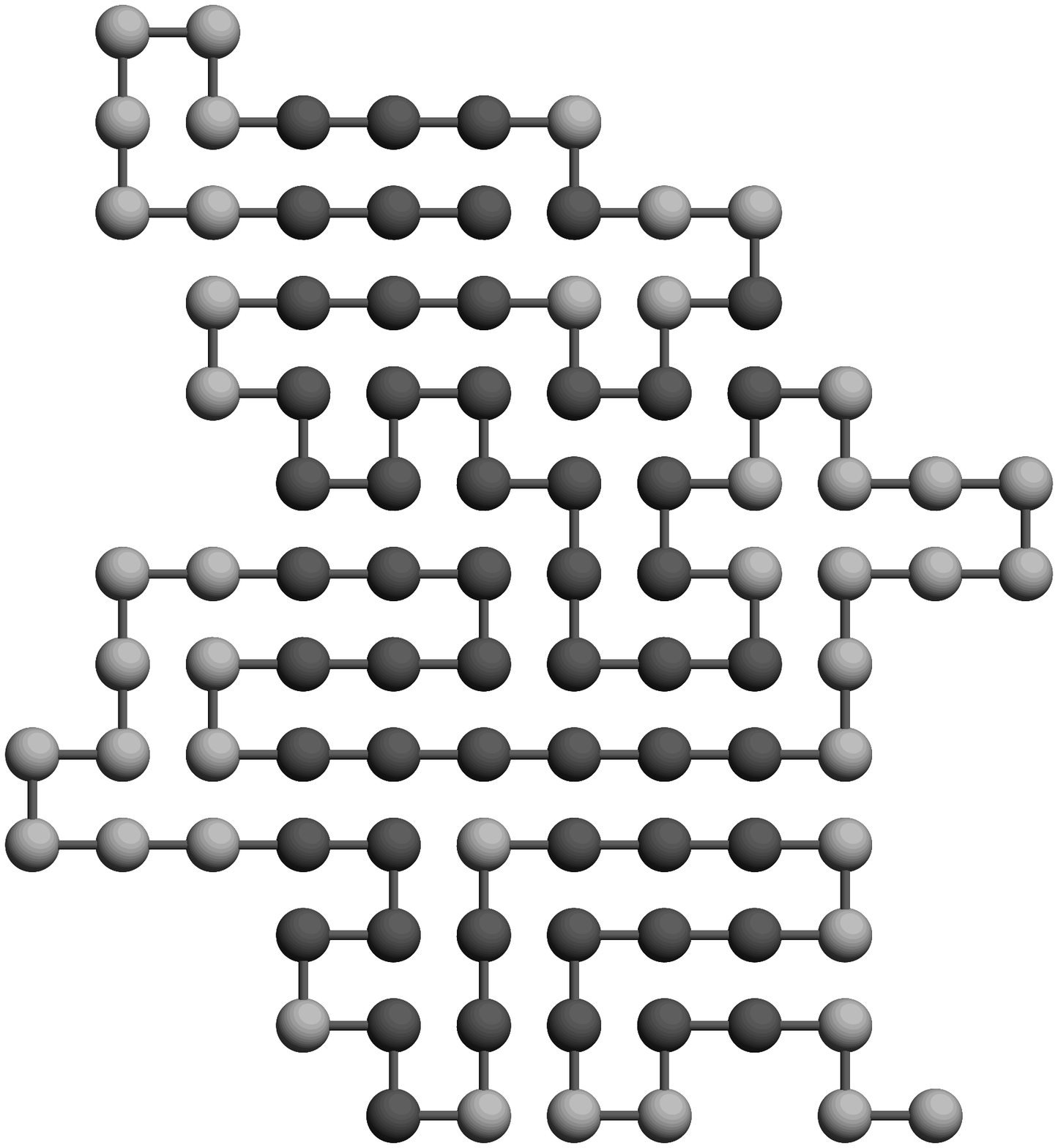,height=7.0cm}
    \end{center}
    {FIG.3. A typical conformation with $E=-50$
     of the second sequence in Table I.
     This value of the energy is reached for the first time by MSOE.}
\label{newly_found_conformation}
\end{figure}

%%%%%%%%%%%%%%%%%%
% 3d beta-helix  %
%%%%%%%%%%%%%%%%%%
Next, we discuss a three dimensional example.
Yue and Dill\cite{Yue&Dill95} have
given a sequence of the chain length $N=42$,
whose ground states are 4-fold degenerate
and their conformations
resemble the parallel $\beta$-helix
found for {\sl Pectate Lyase C}\cite{Yonder}.
The ground state energy has been calculated exactly as $E=-34$
by CHCC method.
As far as we know, no attempt have been reported so far
to calculate themodynamic properties of this sequence.
We appleid MSOE to this sequence.
By taking $E_{min} = -50$ and  $V_{max} = 8$,
we successfully obtained appropriate weight factor
within 50 hours.
Temperature dependence of the specific heat,
the entropy and the gyration radius are shown in Fig.~4.
We see two peaks in the specific heat curve.
The peak at $T\sim 0.28$ and the other one at $T\sim 0.52$
correspond to the transition between
the ground states and compact globule states
(first--order--like) and one between the compact globule states
and the random coil (second--order--like), respectively.
The gyration radius is considerably small at $T\sim 0.52$.
We also found that (not shown in the figure)
the fluctuation of the gyration radius
has a peak at temperature significantly higher than $T\sim 0.52$.
These observations imply that the coil--globule transition
takes place in two steps.
\begin{figure}[ht]
    \begin{center}
    \psfig{file=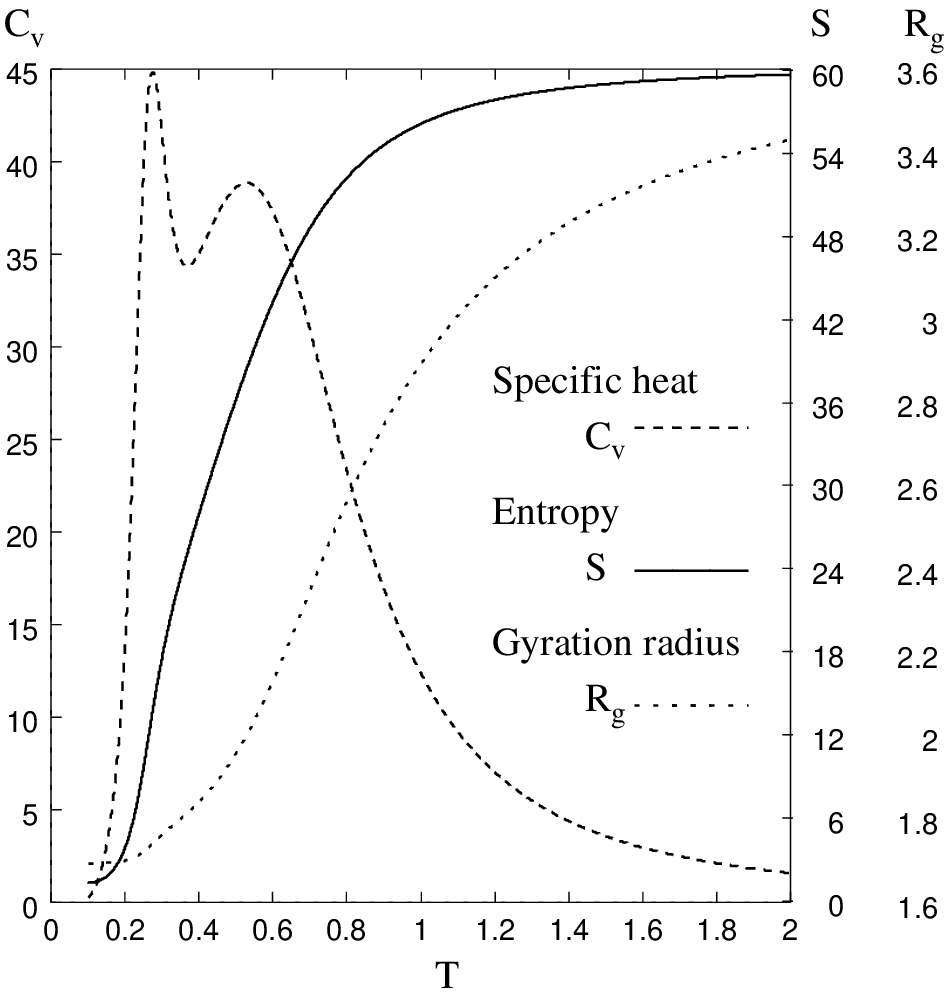,height=6.0cm}
    \end{center}
    {FIG.4.Temperature dependence of
    the specific heat, the entropy, and the gyration radius
    for the $\beta$--helix sequence.}
\end{figure}

%%%%%%%%%%%%
% others   %
%%%%%%%%%%%%
We made MSOE simulations for two more HP sequences:
another sequence of $N=100$ on the square lattice
given in the third row of Table I
and the $N=48$ sequence on the cubic lattice given as the sequence No.9 in ref.\cite{Yue}.
For both sequences, we easily obtained appropreate weight factors
down to the lowest energy states ($E=-47$ and $E=-34$, respectively),
and found that no first--order--like transition takes place.
Thus, not all the HP sequences exhibits first--order--like transition.
The lowest energy states of these sequences are highly degenerated.
On the other hand, the lowest energy states of
2D $N=64$ sequence and the $\beta$--helix sequence,
both of which exhibit first--order--like native--globule transitions
as we have seen above,
have highly regular shapes and have low degeneracy.
The above results imply that
low degeneracy of the ground states is required
for first--order--like native--globule transitions to take place
in long HP chains.

%%%%%%%%%%%%%%%%%%%%%%%%%%%%%%%
% Conventional Multicanonical %
%%%%%%%%%%%%%%%%%%%%%%%%%%%%%%%
We also tested the convensional multicanonical algorithm
for all the sequences studied above,
but none of the lowest energy states found by MSOE
could not be reached within 4-5 days.
Thus, for such long chains as treated in the present study,
the convensional multicanonical algorithm is of no practical use .

%%%%%%%%%%%%%%%%%%%%%%%
%      Summary        %
%%%%%%%%%%%%%%%%%%%%%%%
In summary, we
applied the Multi-Self-Overlap Ensemble (MSOE) method for long chains
of the HP lattice protein model in two and three dimensions.
For all five sequences we treated,
MSOE successfully reached the lowest energy states reported so far.
Especially for the sequence in the second row of Table 1,
we found the lowest energy states
which any other methods have failed to find.
We thus confirmed that MSOE is a powerfull tool for
searching the ground states of lattice protein models.
Moreover, we explored the native--globule transitions at finite
temperature by MSOE,
and found that the degeneracy of the ground
states is relevant for the order of the transition.
MSOE is a quite general method so that it can be applied
to any types of interactions.
It can calculate
correct thermodynamic properties within a moderate CPU time
(although not fastest in some cases actually) as well as low energy  
states.
Especially, once the proper weight factors are determined,
we can calculate thermodynamic properties in
{\it a arbitrary wide range of temperature}
in a single run of MSOE.

\end{multicols}

\end{document}